\documentclass[preprint,aps,showpacs,nofootinbib,preprintnumbers,amsmath,amssymb]{revtex4-1}
\usepackage{}
\usepackage{epsfig}
\usepackage{subfigure}
\usepackage{dcolumn}
\usepackage{bm}
\usepackage[usenames ,dvipsnames]{xcolor}
\usepackage{slashed}
\usepackage{graphicx,color}

\begin{document}
\title{Fragmentation fractions of two-body $b$-baryon decays}

\author{Y.K. Hsiao$^{1,2,3}$, P.Y. Lin$^{4}$, L.W. Luo$^{5}$, and C.Q. Geng$^{1,2,3}$}
\affiliation{
$^{1}$Chongqing University of Posts \& Telecommunications, Chongqing, 400065, China\\
$^{2}$Physics Division, National Center for Theoretical Sciences, Hsinchu, Taiwan 300\\
$^{3}$Department of Physics, National Tsing Hua University, Hsinchu, Taiwan 300\\
$^{4}$Department of Electrophysics, National Chiao Tung University,   Hsinchu, Taiwan 300\\
$^{5}$Institute of Physics, National Chiao Tung University, Hsinchu, Taiwan 300}
\date{\today}

\begin{abstract}
We study the fragmentation fractions $(f_{{\bf B}_b})$ of the $b$-quark to  b-baryons (${\bf  B}_b$).
By the assumption of $f_{\Lambda_b}/(f_u+f_d)=0.25\pm 0.15$ 
in accordance with the  measurements by  LEP, CDF and LHCb Collaborations,
we estimate that $f_{\Lambda_b}=0.175\pm 0.106$ and $f_{\Xi_b^{-,0}}=0.019\pm 0.013$.
From these fragmentation fractions, we derive
${\cal B}(\Lambda_b\to J/\psi \Lambda)=(3.3\pm 2.1)\times 10^{-4}$,
${\cal B}(\Xi_b^-\to J/\psi \Xi^-)=(5.3\pm 3.9)\times 10^{-4}$ and
${\cal B}(\Omega_b^-\to J/\psi \Omega^-)>1.9\times 10^{-5}$.
The predictions of
${\cal B}(\Lambda_b\to J/\psi \Lambda)$ and ${\cal B}(\Xi_b^-\to J/\psi \Xi^-)$
clearly enable us to test the theoretical models,
such as the QCD factorization approach in the $b$-baryon decays.
\end{abstract}

\maketitle
\section{introduction}
The LHCb Collaboration has recently published 
the measurements of the $b$-baryon (${\bf B}_b$) decays~\cite{Aaij:2013pka,Aaij:2014zoa,Aaij:2014pha},
such as the charmful $\Lambda_b$ decays of
$\Lambda_b\to \Lambda_c^+ (K^-,\pi^-)$,
$\Lambda_b\to \Lambda_c^+(D^-,D_s^-)$,
$\Lambda_b\to D^0 p (K^-,\,\pi^-)$, and $\Lambda_b\to J/\psi p(K^-,\pi^-)$,
which are important and interesting results.
For example,
while the $p\pi$ mass distribution 
in $\Lambda_b\to J/\psi p\pi^-$~\cite{Aaij:2014zoa} suggests the existence of the higher-wave baryon,
such as $N(1520)$ or $N(1535)$, a peaking data point in the $Dp$ mass distribution 
in $\Lambda_b\to D^0 p (K^-,\,\pi^-)$~\cite{Aaij:2013pka}
hints  at the resonant $\Sigma_c(2880)$ state.
On the other hand, it is typical to have the partial  observations for
the decay branching ratios,  given by~\cite{pdg}
\begin{eqnarray}\label{Bb_fBb}
{\cal B}(\Lambda_b\to J/\psi \Lambda)f_{\Lambda_b}&=&(5.8\pm 0.8)\times 10^{-5}\,,\nonumber\\
{\cal B}(\Xi_b^-\to J/\psi \Xi^-)f_{\Xi_b^-}&=&(1.02^{+0.26}_{-0.21})\times 10^{-5}\,,\nonumber\\
{\cal B}(\Omega_b^-\to J/\psi \Omega^-)f_{\Omega_b^-}&=&(2.9^{+1.1}_{-0.8})\times 10^{-6}\,,
\end{eqnarray}
where 
$f_{{\bf B}_b}$ are the fragmentation fractions of 
the $b$ quark to $b$-baryons ${\bf B}_b=\Lambda_b$, $\Xi_b^-$ and $\Omega_b^-$.
The partial observations in Eq.~(\ref{Bb_fBb}) along with the measurements of the $\Xi_b^0$ 
decays~\cite{Aaij:2013pka,pdg,Aaij:2014lpa}
are due to the fact that  $f_{\Lambda_b,\Xi_b^{-,0},\Omega_b^-}$
are not well determined.
In the assumption of $f_{\Lambda_b}\simeq f_{baryon}$ 
with $f_{baryon}\equiv {\bf B}(b\to \text{all $b$-baryons})$, it is often adopted that 
$f_{\Lambda_b}=0.1$~\cite{Abdallah:2003gn,Gutsche:2013oea}\footnote{$f_{baryon}\sim 0.1$ was also taken 
 in the previous versions of the PDG.}.
However, according to the recent observations of 
the relatively less decays associated with $\Xi_b^{-,0}$ and $\Omega_b^-$~\cite{HerediaDeLaCruz:2011yi},
$f_{\Lambda_b}\simeq f_{baryon}$ is no longer true. 
As a result, it is urgent to improve the value of $f_{\Lambda_b}$ 
and obtain the less known ones of $f_{\Xi_b^{-,0}}$.
 
Although it is possible to estimate $f_{\Lambda_b}$ by the ratio of $f_{\Lambda_b}/(f_u+f_d)$
with $f_{u,d,s}\equiv{\cal B}(b\to B^-,\bar B^0,\bar B^0_s)$,
different measurements on $f_{\Lambda_b}/(f_u+f_d)$   are not in good agreement,
given by
\begin{eqnarray}
\label{CDF_LEP}
f_{\Lambda_b}/(f_u+f_d)&=&
0.281\pm 0.012({\rm stat})^{+0.058}_{-0.056}({\rm sys})^{+0.128}_{-0.087}({\rm Br})\,~\text{\cite{fLb_CDF}}\,,
\nonumber\\
f_{\Lambda_b}/(f_u+f_d)&=&0.125\pm 0.020\,~\text{\cite{pdg}}\,,
\end{eqnarray}
with the uncertainty related to ${\rm Br}$ due to the uncertainties on the measured branching ratios,
where the first relation given by the CDF Collaboration~\cite{fLb_CDF} is obviously two times larger than 
the world averaged value of the second one~\cite{pdg}, 
dominated by the LEP measurements on Z decays. 
Moreover, since the recent measurements by the LHCb Collaboration also indicate
this inconsistency~\cite{fLb_pT0,fLb_pT,fLb_pT2}, 
it is clear that the values of $f_{\Lambda_b}$ and $f_{\Xi_b^{0,-}}$
can not be experimentally determined yet. 
In this paper, we will demonstrate the possible range for  $f_{\Lambda_b}/(f_u+f_d)$
in accordance with the measurements by LEP, CDF and  LHCb Collaborations
and give
 the theoretical estimations of $f_{\Lambda_b}$ and $f_{\Xi_b^{0,-}}$, 
 which allow us to extract 
${\cal B}(\Lambda_b\to J/\psi \Lambda)$,
${\cal B}(\Xi_b^-\to J/\psi \Xi^-)$, and 
${\cal B}(\Omega_b^-\to J/\psi \Omega^-)$
from the data in Eq.~(\ref{Bb_fBb}). Consequently,
we are able to test the theoretical approach
based on the factorization ansatz, which have been used to calculate 
the two-body ${\bf B}_b$ decays~\cite{Gutsche:2013oea,Liu:2015qfa,Wei:2009np,
Fayyazuddin:1998ap,Chou:2001bn,Ivanov:1997ra,Cheng:1996cs,Ahmed:2011dd}.

\section{Estimations of $f_{\Lambda_b}$ and $f_{\Xi_b^{-,0}}$}
Experimentally, in terms of the specific cases of 
the charmful $\Lambda_b\to\Lambda_c^+\pi^-$ and $\bar B^0\to D^+\pi^-$ decays
or the semileptonic $\Lambda_b\to \Lambda_c^+\mu^- \bar \nu X$ and 
$\bar B\to D\mu^-\bar \nu X$ decays detected with the bins of $p_T$ and $\eta$, 
where $p_T$ is the transverse momentum and $\eta=-\ln(\tan\theta/2)$ is the pseudorapidity 
defined by the polar angle $\theta$ 
with respect to the beam direction~\cite{fLb_CDF,fLb_pT0,fLb_pT},
the ratio of $f_{\Lambda_b}/(f_u+f_d)$ can be related to $p_T$ and $\eta$. 
This explains the inconsistency between the results from CDF and LEP with $p_T=15$ and 45 GeV, respectively.
While $f_s/f_u$ is measured with  slightly dependences on $p_T$ and $\eta$~\cite{Aaij:2013qqa},
 $f_{\Lambda_b}/(f_u+f_d)$ is 
fitted as the linear form in Ref.~\cite{fLb_pT0} with $p_T=0-14$ GeV and 
the exponential form in Refs.~\cite{fLb_pT,fLb_pT2} with $p_T=0-50$ GeV, respectively,
for the certain  range of $\eta$.

\subsection{The present status of  $f_{\Lambda_b}/(f_u+f_d)$} 
With the semileptonic $\Lambda_b\to \Lambda_c^+\mu^- \bar \nu X$ and 
$\bar B\to D\mu^-\bar \nu X$ decays, the LHCb Collaboration has shown 
the dependence of $f_{\Lambda_b}/(f_u+f_d)$ on $p_T$ in the range of $p_T=0-14$ GeV
to be the linear form, given by~\cite{fLb_pT}
\begin{eqnarray}\label{fLb_14}
f_{\Lambda_b}/(f_u+f_d)&=&\big(0.404\pm0.017({\rm stat}) \pm0.027({\rm syst}) \pm0.105({\rm Br})\big)
\nonumber\\
&&\big(1-[0.031\pm0.004({\rm stat})\pm0.003({\rm syst})]p_T\big)\,,
\end{eqnarray} 
where ${\rm Br}$ arises from the absolute scale 
uncertainty due to the poorly known branching ratio of ${\cal B}(\Lambda_c^+\to pK^-\pi^+$).
By averaging $f_{\Lambda_b}/(f_u+f_d)$ with $p_T=0-14$ GeV, we obtain
\begin{eqnarray}\label{averagedfLb}
\bar f_{\Lambda_b}=(0.316\pm 0.087)(f_u+f_d)\,,
\end{eqnarray}
which agrees with the first relation in Eq.~(\ref{CDF_LEP}) given by the CDF Collaboration
with $p_T\simeq$15 GeV. On the other hand, 
with the charmful $\Lambda_b\to\Lambda_c^+\pi^-$ and $\bar B^0\to D^+\pi^-$ decays,
another analysis by the LHCb Collaboration presents
the exponential dependence of $f_{\Lambda_b}/f_d$ on $p_T$~\cite{fLb_pT,fLb_pT2}: 
\begin{eqnarray}\label{fLb_50}
f_{\Lambda_b}/f_d&=&(0.151\pm 0.030)+exp\{-(0.57\pm 0.11)-(0.095\pm 0.016)p_T\}\,,
\end{eqnarray} 
with the wider range of $p_T= 0-50$ GeV. 
By averaging the value in Eq.~(\ref{fLb_50}) with $p_T=0-50$ GeV, we find
\begin{eqnarray}\label{averagedfLb2}
\bar f_{\Lambda_b}=(0.269\pm 0.040)f_d=(0.135\pm 0.020)(f_u+f_d)\,,
\end{eqnarray}
with $f_u=f_d$ due to the isospin symmetry, 
where the error has combined the uncertainties in Eq. (\ref{fLb_50}).
It is interesting to note that, as the relation in Eq.~(\ref{fLb_50}) with $p_T=0-50$ GeV
overlaps $p_T\simeq 45$ GeV for the second relation from LEP in Eq.~(\ref{CDF_LEP}), 
its value of $\bar f_{\Lambda_b}=(0.135\pm 0.020)(f_u+f_d)$ is close to
the LEP result of $f_{\Lambda_b}=(0.125\pm 0.020)(f_u+f_d)$.
Apart form the values in Eqs.~(\ref{averagedfLb}) and (\ref{averagedfLb2}),
the reanalyzed results by  CDF and LHCb Collaborations give
$f_{\Lambda_b}/(f_u+f_d)$ to be $0.212\pm 0.058$ and $0.223\pm 0.022$
with the averaged $p_T\simeq 13$ and 7 GeV, respectively~\cite{fLb_pT2}.
We hence make the assumption of 
\begin{eqnarray}\label{RLb}
R_{\Lambda_b}\equiv f_{\Lambda_b}/(f_u+f_d)=0.25\pm 0.15\,,
\end{eqnarray}
to cover the possible range in accordance with the measurements from
the three  Collaborations of LEP, CDF and LHCb, which will be used 
to  estimate the values of $f_{\Lambda_b}$ and $f_{\Xi_b^{0,-}}$
in the following.

\subsection{Theoretical determination of $f_{\Xi_b^-}/f_{\Lambda_b}$}
In principle, when the ratios of $f_{\Lambda_b}/(f_u+f_d)$ and 
$f_{\Xi_b^{0,-}}/f_{\Lambda_b}$ are both known,
by adding the relations of
~\cite{pdg,Aaij:2013qqa}
\begin{eqnarray}\label{3relations}
&&f_u+f_d+f_s+f_{baryon}=1\,,\nonumber\\
&&f_{baryon}\simeq f_{\Lambda_b}+f_{\Xi_b^-}+f_{\Xi_b^0}\,,\nonumber\\
&&f_s=(0.256\pm 0.020)f_d\,,
\end{eqnarray}
and  $f_u=f_d$ as well as $f_{\Xi_b^-}=f_{\Xi_b^0}$ due to the isospin symmetry, 
we can derive the values of $f_u$, $f_d$, $f_s$, $f_{\Lambda_b}$, $f_{\Xi_b^-}$ and $f_{\Xi_b^0}$.
For $f_{\Xi_b^{-}}/f_{\Lambda_b}$, 
it was once given that
\begin{eqnarray}\label{firstXiLb}
&&f_{\Xi_b^{-}}/f_{\Lambda_b}\simeq f_s/f_u\,~\text{\cite{HerediaDeLaCruz:2011yi,Abazov:2007am}}\,,
\nonumber\\
&&f_{\Xi_b^{0}}/f_{\Lambda_b}\simeq 0.2\,~\text{\cite{PRL113}}\,,
\end{eqnarray}
where the first relation from Refs.~\cite{HerediaDeLaCruz:2011yi,Abazov:2007am} 
requires the assumption of 
$R_1\equiv{\cal B}(\Xi_b^-\to J/\psi \Xi^-)/$ ${\cal B}(\Lambda_b\to J/\psi \Lambda)\simeq 1$~\cite{fLb_pT},
while the second one from Ref.~\cite{PRL113} uses
$R_2\equiv{\cal B}(\Xi_b^0\to \Xi_c^+ \pi^-)/{\cal B}(\Lambda_b\to \Lambda_c^+\pi^-)\simeq 1$
along with $R_3\equiv{\cal B}(\Xi_c^+\to p K^- \pi^+)/{\cal B}(\Lambda_c^+\to p K^-\pi^+)\simeq 0.1$
from the naive Cabibbo factors.
However, we note that
the theoretical calculations provide us with more understanding of $b$-baryon decays,
such as the difference between the $\Lambda_b\to \Lambda$ and $\Xi_b^-\to \Xi^-$ transitions,
 based on the $SU(3)$ flavor and $SU(2)$ spin symmetries.
As a result, the assumption of $R_1=R_2\simeq 1$ might be too naive.
Since the theoretical approach with the factorization ansatz
well explains ${\cal B}(\Lambda_b\to p \pi^-)$ and ${\cal B}(\Lambda_b\to p K^-)$,
and particularly the ratio of ${\cal B}(\Lambda_b\to p \pi^-)/{\cal B}(\Lambda_b\to p K^-)\sim 0.84$~\cite{Hsiao:2014mua}, 
it can be reliable to determine $f_{\Xi_b^-}/f_{\Lambda_b}$.

Theoretically, we use the factorization approach to calculate 
the two-body $b$-baryon decay, such that the amplitude  corresponds to
the decaying process of the ${\bf B}_b\to {\bf B}_n$ transition with the recoiled meson. 
\begin{figure}[t]
\centering
\includegraphics[width=2.8in]{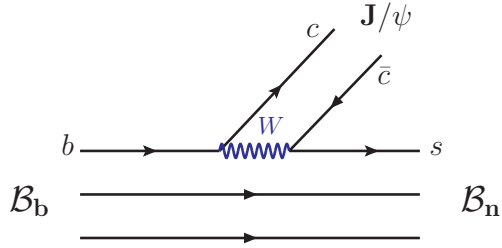}
\caption{The ${\bf B}_b\to {\bf B}_n J/\psi$ decays via
the internal $W$-boson emission diagram.}\label{BbtoBnMc}
\end{figure}
Explicitly, as shown in Fig.~\ref{BbtoBnMc}, where the $W$-boson emission is internal,
the amplitude via the quark-level $b\to c\bar c s$ transition can be factorized as
\begin{eqnarray}\label{eq2}
{\cal A}({\bf B}_b\to {\bf B}_n J/\psi)&=&
\frac{G_F}{\sqrt 2}V_{cb}V_{cs}^*a_2\,
\langle J/\psi|\bar c\gamma^\mu(1- \gamma_5) c|0\rangle
\langle {\bf B}_n|\bar s\gamma_\mu(1-\gamma_5) b|{\bf B}_b\rangle\,, 
\end{eqnarray}
for  $\Lambda_b\to \Lambda J/\psi$ or $\Xi_b^-\to \Xi^-J/\psi$,
where the parameter $a_2$ is given by~\cite{Hamiltonian,ali}
\begin{eqnarray}\label{a12}
a_2=c_2^{eff}+\frac{c_1^{eff}}{N_c}\,,
\end{eqnarray}
with the effective Wilson coefficients $(c^{eff}_1,\,c^{eff}_2)=(1.168,-0.365)$.
Note that the color number $N_c$ originally being equal to 3 in the naive factorization,
which gives $a_2=0.024$ in Eq.~(\ref{a12}),
should be taken as a floating number from $2\to\infty$
to account for the non-factorizable effects in the generalized factorization.
The matrix element for the $J/\psi$ production is given by
$\langle J/\psi|\bar c \gamma_\mu c|0\rangle=m_{J/\psi} f_{J/\psi}\varepsilon_\mu^*$
with $m_{J/\psi}$, $f_{J/\psi}$, and $\varepsilon_\mu^*$ as 
the mass,  decay constant, and  polarization vector, respectively.
The matrix elements of the ${\bf B}_b\to {\bf B}_n$ baryon
transition in Eq.~(\ref{eq2}) have the general forms:
\begin{eqnarray}\label{FFs}
&&\langle {\bf B}_n|\bar q \gamma_\mu b|{\bf B}_b\rangle=
\bar u_{{\bf B}_n}[f_1\gamma_\mu+\frac{f_2}{m_{{\bf B}_b}}i\sigma_{\mu\nu}q^\nu+
\frac{f_3}{m_{{\bf B}_b}}q_\mu] u_{{\bf B}_b}\,,\nonumber\\
&&\langle {\bf B}_n|\bar q \gamma_\mu\gamma_5 b|{\bf B}_b\rangle=
\bar u_{{\bf B}_n}[g_1\gamma_\mu+\frac{g_2}{m_{{\bf B}_b}}i\sigma_{\mu\nu}q^\nu+
\frac{g_3}{m_{{\bf B}_b}}q_\mu]\gamma_5 u_{{\bf B}_b}\,,
\end{eqnarray}
where $f_j$ ($g_j$) ($j=1,2,3$)  are the form factors,
with $f_{2,3}=0$ and $g_{2,3}=0$ due to the helicity conservation,
as derived in Refs.~\cite{CF,Wei:2009np,Gutsche:2013oea}.
It is interesting to note that, as the helicity-flip terms,
the theoretical calculations from the loop contributions
to $f_{2,3}$ ($g_{2,3}$) indeed result in the values
to be one order of magnitude smaller than that to $f_1(g_1)$,
which can be safely neglected.
In the double-pole momentum dependences, 
one has that~\cite{Hsiao:2014mua}
\begin{eqnarray}
F(q^2)=\frac{F(0)}{(1-q^2/m_{{\bf B}_b}^2)^2}\,,\;\; (F=f_1\,,~g_1).
\end{eqnarray}
We are able to relate different ${\bf B}_b\to {\bf B}_n$ transition form factors
based on  $SU(3)$ flavor and  $SU(2)$ spin symmetries,
which have been used to connect the space-like ${\bf B}_n\to {\bf B}'_n$ transition form factors
in the neutron decays~\cite{Brodsky1}, and the time-like $0\to {\bf B}_n \bar {\bf B}'_n$ baryonic 
as well as $B\to {\bf B}_n \bar {\bf B}'_n$ transition form factors
in the baryonic $B$ decays~\cite{Chua:2002wn,Chua:2002yd,Chen:2008sw,Geng:2011pw,Hsiao:2014zza}.
As a result, we obtain $(f_1(0),g_1(0))=(C,C)$, $(-\sqrt{2/3}C, -\sqrt{2/3}C)$, and $(0,0)$ with $C$ a constant
for $\langle p|\bar u\gamma_\mu(\gamma_5)b|\Lambda_b \rangle$,
$\langle \Lambda|\bar s\gamma_\mu(\gamma_5)b|\Lambda_b \rangle$, and 
$\langle \Sigma^0|\bar s\gamma_\mu(\gamma_5)b|\Lambda_b \rangle$, 
which are the same as those 
in Ref.~{\cite{CF}} based on 
the heavy-quark and large-energy symmetries 
for the $\Lambda_b\to (p,\Lambda,\Sigma^0)$ transitions, respectively. 
In addition, we have 
$f_1(0)=g_1(0)=C$ for $\langle \Xi^-|\bar s\gamma_\mu(\gamma_5)b|\Xi_b^- \rangle$.
To obtain the branching ratio for the two-body decays, the equation
is given by~\cite{pdg}
\begin{eqnarray}
{\cal B}({\bf B}_b\to J/\psi {\bf B}_n)=
\frac{\Gamma({\bf B}_b\to J/\psi {\bf B}_n)\tau_{{\bf B}_b}}{6.582\times 10^{-25}}\,,
\end{eqnarray}
with $\tau_{{\bf B}_b}$ the life time, where 
\begin{eqnarray}
\Gamma({\bf B}_b\to J/\psi {\bf B}_n)=\frac{|\vec{P}_{J/\psi}|}{8\pi m_{{\bf B}_b}^2}
|{\cal A}({\bf B}_b\to J/\psi {\bf B}_n)|^2\,,
\end{eqnarray}
with $|\vec{P}_{J/\psi}|=|\vec{P}_{{\bf B}_n}|=
\{[m_{{\bf B}_b}^2-(m_{J/\psi}+m_{{\bf B}_n})^2][m_{{\bf B}_b}^2-(m_{J/\psi}-m_{{\bf B}_n})^2]\}^{1/2}/
(2m_{{\bf B}_b})$. As a result, we obtain
\begin{eqnarray}\label{data2}
\frac{{\cal B}(\Xi_b^-\to J/\psi \Xi^-)}{{\cal B}(\Lambda_b\to J/\psi \Lambda)}=
\frac{\tau_{\Xi_b^-}}{\tau_{\Lambda_b}}
\frac{C^2}{(-\sqrt{2/3}C)^2}=1.63\pm 0.04\,,
\end{eqnarray}
with ${\tau_{\Xi_b^-}}/{\tau_{\Lambda_b}}=1.089\pm 0.026\pm 0.011$~\cite{Aaij:2014lxa}.
We note that, theoretically, $R_1=1.63$ apparently  deviates by 63\% from $R_1=1$ in the simple assumption.
To determine $f_{\Xi_b^-}/f_{\Lambda_b}$, we relate Eq.~(\ref{data2}) to (\ref{Bb_fBb}) to give
\begin{eqnarray}\label{newXiLb}
f_{\Xi_b^-}=(0.108\pm 0.034)f_{\Lambda_b}\,,
\end{eqnarray}
which is different from the numbers in Eq.~(\ref{firstXiLb}).

\subsection{Determinations of $f_{\Xi_b^{-,0}}$ and $f_{\Lambda_b}$}
According to Eqs.~(\ref{averagedfLb}), (\ref{RLb}), (\ref{3relations}) and (\ref{newXiLb}), we  
derive the values of $f_u$, $f_d$, $f_s$, $f_{\Lambda_b}$, $f_{\Xi_b^-}$ and $f_{\Xi_b^0}$
in Table~\ref{fLb_R}, which agree with the data in the PDG~\cite{pdg}.
Note that $f_{\Omega_b^-}<0.108$ is from the error in $f_{baryon}$.
In addtion, 
 $f_{baryon}=0.213\pm 0.108$,
  which overlaps $0.089\pm 0.015$ from Z-decays~\cite{pdg}  and $0.237\pm 0.067$
from Tevatron~\cite{pdg}, is  due to
the assumption of $R_{\Lambda_b}=0.25\pm 0.15$ in Eq.~(\ref{RLb}) 
to cover the possible range from the data.
Similarly,  $f_{\Lambda_b}=0.175\pm 0.106$ overlaps 
 $f_{\Lambda_b}=0.07$ from the LEP measurements~\cite{Galanti:2015pqa},
while  $f_{\Xi_b^-}=f_{\Xi_b^0}=0.019\pm 0.013$ is consistent with  
 $f_{\Xi_b^-}=0.011\pm 0.005$ from the measurement~\cite{fXib_LEP}.
\begin{table}[b]
\caption{Results of  $f_i$ 
($i=u$, $d$, $s$, baryon, $\Lambda_b$, $\Xi_b^{-,0}$, and $\Omega_b^-$),
compared with those from Z-decays and Tevatron in PDG~\cite{pdg}.}
\label{fLb_R}
\begin{tabular}{|c|ccc|}
\hline
& our result
& Z-decays~\cite{pdg}& Tevatron~\cite{pdg}\\\hline
$f_u=f_d$  &$0.349\pm 0.037$  &$0.404\pm 0.009$ &$0.330\pm 0.030$\\
$f_s$  &$0.089\pm 0.018$  &$0.103\pm 0.009$ &$0.103\pm 0.012$\\
%
%
$f_{baryon}$  &$0.213\pm 0.108$  &$0.089\pm 0.015$ &$0.237\pm 0.067$\\
$f_{\Lambda_b}$  &$0.175\pm 0.106$  &--- &---\\
$f_{\Xi_b^-}=f_{\Xi_b^0}$  &$0.019\pm 0.013$  &--- &---\\
$f_{\Omega_b^-}$  &$<0.108$  &--- &---
\\\hline
\end{tabular}
\end{table}
We hence convert the data in Eq.~(\ref{Bb_fBb}) to be
\begin{eqnarray}\label{Bb_f}
{\cal B}(\Lambda_b\to J/\psi \Lambda)&=&(3.3\pm 2.1)\times 10^{-4}\,,\nonumber\\
{\cal B}(\Xi_b^-\to J/\psi \Xi^-)&=&(5.3\pm 3.9)\times 10^{-4}\,,\nonumber\\
{\cal B}(\Omega_b^-\to J/\psi \Omega^-)&>&1.9\times 10^{-5}\,,
\end{eqnarray}
with ${\cal B}(\Xi_b^-\to J/\psi \Xi^-)\simeq 1.6{\cal B}(\Lambda_b\to J/\psi \Lambda)$
to be in accordance with Eq.~(\ref{data2}).
With the use of $f_{\Xi_b^{0,-}}$ ,we can also estimate the $\Xi_b^{0,-}$ decays ~\cite{pdg,Aaij:2014lpa},
given by
\begin{eqnarray}
{\cal B}(\Xi_b^-\to \Xi^- \ell^- \bar \nu_\ell X)&=&(2.1\pm 1.5)\times 10^{-2}\,,\nonumber\\
{\cal B}(\Xi_b^0\to \bar K^0 p\pi^-)&=&(1.1\pm 1.5)\times10^{-5}\,,\nonumber\\
{\cal B}(\Xi_b^0\to \bar K^0 p K^-)&=&(1.1\pm 1.1)\times10^{-5}\,,\nonumber\\
{\cal B}(\Xi_b^0\to D^0 p K^-)&=&(9.5\pm 9.4)\times 10^{-5}\,,\nonumber\\
{\cal B}(\Xi_b^0\to \Lambda_c^+ K^-)&=&(4.2\pm 4.7)\times 10^{-5}\,.
\end{eqnarray}

\subsection{Test of the non-factorizable effects}
To numerically test the non-factorizable effects, 
the CKM matrix elements in the Wolfenstein parameterization 
are taken as $(V_{cb},V_{cs})=(A\lambda^2,1-\lambda^2/2)$
with $(\lambda,\,A)=(0.225,\,0.814)$~\cite{pdg}, while
$f_{J/\psi}=418\pm 9$ MeV~\cite{Becirevic:2013bsa}.
 The constant value of $C$ in Ref.~\cite{Hsiao:2014mua} is fitted to be $C=0.136\pm 0.009$ 
to explain the branching ratios and predict the CP violating asymmetries of $\Lambda_b\to p(K^-,\pi^-)$, 
which is also consistent with the value of  $0.14\pm 0.03$ 
in the  light-cone sum rules~\cite{CF} and 
those in Refs.~\cite{Wei:2009np,Gutsche:2013oea}.

To explain the branching ratios of 
$\Lambda_b\to J/\psi \Lambda$ and $\Xi_b^-\to J/\psi \Xi^-$ in Eq.~(\ref{Bb_f}),
the floating color number $N_c$ is evaluated to be
\begin{eqnarray}\label{Nc}
N_c=2.15\pm 0.17\,,
\end{eqnarray}
which corresponds to $a_2= 0.18\pm 0.04$, 
in comparison with $a_2=0.024$ for $N_c=3$. 
Note that since $N_c=2.15$ in Eq.~(\ref{Nc}) is not far from  3,
we conclude that the non-factorizable effects are controllable.
As a result, the theoretical approach based on the factorization ansatz
is demonstrated to be reliable to explain the two-body ${\bf B}_b$ decays.

\section{Conclusions}

In sum, we made the assumption of $R_{\Lambda_b}=f_{\Lambda_b}/(f_u+f_d)=0.25\pm 0.15$,
which is in accordance with the the measurements by  LEP, CDF and LHCb Collaborations.
We have estimated that 
$f_{\Lambda_b}=0.175\pm 0.106$ and $f_{\Xi_b^{-,0}}=0.019\pm 0.013$,
which can be used to extract the branching ratios
of the anti-triplet $b$-baryon decays. 
Explicitly, we have found
${\cal B}(\Lambda_b\to J/\psi \Lambda)=(3.3\pm 2.1)\times 10^{-4}$,
${\cal B}(\Xi_b^-\to J/\psi \Xi^-)=(5.3\pm 3.9)\times 10^{-4}$ and 
${\cal B}(\Omega_b^-\to J/\psi \Omega^-)>1.9\times 10^{-5}$.
We have also demonstrated that the predictions of
${\cal B}(\Lambda_b\to J/\psi \Lambda)$ and ${\cal B}(\Xi_b^-\to J/\psi \Xi^-)$
would help us to test the theoretical models, 
such as the factorization approach.

\section*{ACKNOWLEDGMENTS}
This work was partially supported by National Center for Theoretical
Sciences,  National Science Council
(NSC-101-2112-M-007-006-MY3) and (NSC 101-2112-M-009-004-MY3), 
MoST (MoST-104-2112-M-007-003-MY3)
and National Tsing Hua University~(104N2724E1).

\end{document}